# Polaron Model of the Formation of Hydrated Electron States

**V. D. Lakhno[a], A. V. Volokhova[b], E. V. Zemlyanaya[b], I. V. Amirkhanov[b], I. V. Puzynin[b], and T. P. Puzynina[b]**

[a] *Institute of Mathematical Problems of Biology, Russian Academy of Sciences, Pushchino, Moscow oblast, 142290 Russia*
*e-mail: lak@impb.psn.ru*

[b] *Laboratory of Informational Technologies, Joint Institute for Nuclear Research, Dubna, Moscow oblast, 141980 Russia*
*e-mail: volokhova@jinr.ru, elena@jinr.ru, camir@jinr.ru, ipuzynin@jinr.ru, puzynina@jinr.ru*

**Abstract**—A computer simulation of the formation of photoexcited electrons in water is performed within the framework of a dynamic model. The obtained results are discussed in comparison with experimental data and theoretical estimates.

## INTRODUCTION

In this paper, we present the results of computer simulation of the formation process of photoexcited (solvated, hydrated) electrons in water. In accordance with modern ideas, as a result of the radiolysis of liquid and glassy systems, as a rule, bound (solvated) states of electrons appear; electrons do not form any chemical bonds in them [1–4]. Such states can be described as the capture of an excess electron by a potential well formed as a result of the polarization of medium molecules by the electron.

Interest in the study of solvated electrons is first related to the fact that an enormous number of reactions occur with their participation in nonorganic and organic chemistry. Hydrated electrons, which are the strongest reducer in water, are of special interest. In biological systems, the hydrated electron plays a great role in processes of long-distance charge transfer. The dynamics of the absorption spectra during hydrated-electron formation in irradiated water gives important information about its structure and the kinetics of chemical reactions with its participation. Modern femtosecond spectroscopic methods give detailed information on the dynamics of the formation of hydrated electrons excited in pure water by means of its photoirradiation. The aim of this paper is to develop a calculation method which describes such dynamics adequately.

In section 1, we describe a physical model of a polaron. In section 2, we present the mathematical problem statement and the numerical approach is described. In section 3, we present the results of methodical calculations carried out to verify the correctness of the model its C++ realization and discuss the results of numerical simulation compared with theoretical and experimental estimates. The main conclusions are presented in Conclusions.

## PHYSICAL MODEL

When describing the dynamics of the absorption spectra of hydrated electrons, we proceed from a polaron model in which the polar medium is regarded as continuous. As shown in [5], this assumption is, probably, fulfilled well if the initial state of the excited electron has a rather large size. In this case, to calculate the dynamics of hydrated-electron formation, we proceed from the total energy functional $E$ for the optical Pekar polaron [6, 7]:

$$E = \int d^3r \left\{ \frac{\hbar^2}{2m}|\nabla\Psi|^2 - \mathrm{e}\Phi|\Psi|^2 - \frac{1}{8\pi c\omega^2}(\nabla\dot{\Phi})^2 + \frac{1}{8\pi c}(\nabla\Phi)^2 \right\}, \quad (1)$$

where $m$ is the effective electron mass, $c = \varepsilon_\infty^{-1} - \varepsilon_0^{-1}$, $\varepsilon_\infty$ and $\varepsilon_0$ are the high-frequency and steady-state permittivities, e is the elementary charge, $\omega$ is the frequency of optical polarization medium oscillations, and $\Phi$ is the potential appearing during the polarization process.

The equations of motion corresponding to functional (1) have the forms

$$i\hbar\dot{\Psi}(\mathbf{r}, t) = -\frac{\hbar^2}{2m}\Delta\Psi(\mathbf{r}, t) - \mathrm{e}\Phi(\mathbf{r}, t)\Psi(\mathbf{r}, t), \quad (2)$$

$$\Delta\Phi(\mathbf{r}, t) = \theta(\mathbf{r}, t), \quad (3)$$

$$\frac{1}{4\pi\omega^2 c}\ddot{\theta}(\mathbf{r}, t) + \frac{1}{4\pi c}\theta(\mathbf{r}, t) + \mathrm{e}|\Psi(\mathbf{r}, t)|^2 = 0. \quad (4)$$

In accordance with [5], for the polarization potential $\Phi$, we introduce relaxation (friction) determined by the coefficient $\gamma$ related to the dielectric relaxation time $\tau$ by the expression

$$\gamma^{-1} = \omega^2\tau c. \tag{5}$$

Then Eq. (4) becomes

$$\frac{1}{4\pi\omega^2 c}\ddot{\theta}(\mathbf{r}, t) + \gamma\dot{\theta}(\mathbf{r}, t) + \frac{1}{4\pi c}\theta(\mathbf{r}, t) + \mathrm{e}|\Psi(\mathbf{r}, t)|^2 = 0. \tag{6}$$

Equations (2), (3), and (6) determine the three-dimensional dynamic polaron model.

We note that if the relation $\nabla\Phi = 4\pi\mathbf{P}$, where $\mathbf{P}$ is the medium polarization, is taken into account, Eq. (6) can be represented in the form

$$\ddot{\mathbf{P}} + \frac{1}{\tau}\dot{\mathbf{P}} + \omega^2\mathbf{P} - \frac{\omega^2 c}{4\pi}\mathbf{D} = 0, \tag{7}$$

$$\mathbf{D} = \mathrm{e}\int|\Psi(\mathbf{r}, t)|^2\frac{(\mathbf{r} - \mathbf{r}')}{|\mathbf{r} - \mathbf{r}'|^3}d^3r', \tag{8}$$

where $\mathbf{D}$ has the physical meaning of induction produced by an electron in the polaron medium.

## MATHEMATICAL STATEMENT OF THE PROBLEM AND THE METHOD OF NUMERICAL SOLUTION

Substituting the expansions of the functions $\Psi(\mathbf{r}, t)$, and $\Phi(\mathbf{r}, t)$, $\theta(\mathbf{r}, t)$ in terms of the spherical harmonics $Y_{lm_s}(\theta_s, \varphi_s)$

$$\Psi(\mathbf{r}, t) = \sum_{l=0}^{\infty}\sum_{m_s=-l}^{l}\frac{\psi_{lm_s}(r, t)}{r}Y_{lm_s}(\theta_s, \varphi_s),$$

$$\Phi(\mathbf{r}, t) = \sum_{l=0}^{\infty}\sum_{m_s=-l}^{l}\frac{\varphi_{lm_s}(r, t)}{r}Y_{lm_s}(\theta_s, \varphi_s),$$

$$\theta(\mathbf{r}, t) = \sum_{l=0}^{\infty}\sum_{m_s=-l}^{l}\frac{\theta_{lm_s}(r, t)}{r}Y_{lm_s}(\theta_s, \varphi_s)$$

into (2), (3), and (6) and restricting our consideration to the spherically symmetric case of $m_s = l = 0$, we obtain the system of spatially one-dimensional partial differential equations for functions $\psi_{00}(r, t)$, $\varphi_{00}(r, t)$, and $\theta_{00}(r, t)$, which acquires the following form when passing to dimensionless quantities (here and hereafter, the subscripts 00 are omitted):

$$\begin{cases} \left[i2\bar{m}\frac{\partial}{\partial t} + \frac{\partial^2}{\partial x^2} + 2\bar{m}\frac{r_{00}}{\tilde{\varepsilon}}\frac{\varphi(x, t)}{x}\right]\psi(x, t) = 0, \\ \frac{\partial^2}{\partial x^2}\varphi(x, t) = \theta(x, t), \\ \left[\frac{\partial^2}{\partial t^2} + \bar{\gamma}\frac{\partial}{\partial t} + \bar{\omega}^2\right]\theta(x, t) = -\bar{\omega}^2\frac{|\psi(x, t)|^2}{x}, \end{cases} \tag{9}$$

with boundary conditions

$$\varphi(0, t) = \varphi'(\infty, t) = 0, \ \psi(0, t) = \psi(\infty, t) = 0, \quad \theta(0, t) = \theta(\infty, t) = 0. \tag{10}$$

Equations (9) with boundary conditions (10) describe the evolution of the state given at the initial instant of time. Here, $\tilde{\varepsilon} = 1.81$ is the permittivity coefficient; $\bar{m} = 2.692$, $\bar{\gamma} = 2.145$, and $\bar{\omega} = 1$ are the parameters of the effective mass of the hydrated electron, the relaxation (friction), and the frequency reduced to the dimensionless form; $r_{00}$ is the scaling factor

$$r_{00} = \sqrt{t_{00}} = \sqrt{t_0/t_{A0}} = 164.64;$$

$t_0 = 1/\omega_0$, $\omega_0 = 1.5246 \times 10^{12}$ s$^{-1}$ is the characteristic frequency of the medium oscillations, and $t_{A0} = 2.42 \times 10^{-17}$ is the atomic unit of time. Scaling is introduced to equalize the initial values of the physical parameters which differ by ten orders or more. In this case, the integration interval over the spatial variable decreases significantly (in accordance with the law $x = r/r_{00}$), which simplifies computer simulation.

The transition from the dimensionless variables $t$ and $x$ to dimensional $t_{\text{size}}$ and $r$ is performed using the formulas $t_{\text{size}} = tt_0$ and $r = xr_0$, where $r_0 = r_{00}a$, $a = 0.529 \times 10^{-8}$ cm is the Bohr radius. The relation between the effective mass $m$ of the hydrated electron, the frequency $\omega$ of optical polarization medium oscillations, and the friction coefficient $\gamma$ and the parameters $\bar{m}$, $\bar{\omega}$, and $\bar{\gamma}$ in system (9) is determined by the expressions $m = \bar{m}m_e$ (where $m_e$ is the electron mass in vacuum), $\omega = \bar{\omega}/t_0$, and $\gamma = \bar{\gamma}/t_0$.

The energy integral is calculated in accordance with the formula

$$W(t) = \frac{1}{2\bar{m}}\int\left|\frac{\partial\psi(x, t)}{\partial x}\right|^2 dx - \frac{r_{00}}{\tilde{\varepsilon}}\int\frac{\varphi(x, t)|\psi(x, t)|^2}{x}dx. \tag{11}$$

For numerical solution, system of differential equations (9) is replaced with that of difference equations on a uniform discrete mesh. As a result of substitutions of well-known finite-difference formulas [8], we obtain a system of difference equations; to solve this system, we used the algorithm described in [9] in detail making it possible to successively calculate $\Theta(x, t^n)$, $\varphi(x, t^n)$, and $\psi(x, t^n)$ at nodes of the discrete mesh along the $x$ axis under given initial conditions in each time layer $t^n$. To speed up performance of the calculations, we developed the parallel realization of this algorithm on the basis of MPI technology providing an almost twofold decrease in the calculation time on two-processor (binuclear) systems.

## NUMERICAL RESULTS

### *Steady-State Solutions*

System (9) has steady-state (i.e., time-independent) solutions satisfying the problem of finding the eigenvalues for the system of ordinary differential equations

$$\begin{cases} \left[\dfrac{d^2}{dx^2} - 2\bar{m}\lambda + 2\bar{m}\dfrac{r_{00}}{\bar{\varepsilon}}\dfrac{\Phi_{st}(x)}{x}\right]\Psi_{st}(x) = 0, & 0 \le x \le \infty, \\ \dfrac{d^2}{dx^2}\Phi_{st}(x) = -\dfrac{\Psi_{st}^2(x)}{x}, \end{cases} \quad (12)$$

with the boundary and normalization conditions

$$\begin{cases} \Psi_{st}(0) = 0, \Phi_{st}(0) = 0, \\ \Psi_{st}(\infty) = 0, \Phi'_{st}(\infty) = 0, \end{cases} \quad \int_0^\infty \Psi_{st}^2(x)dx = 1. \quad (13)$$

It can be shown that, if the initial conditions for system (9) in the following form are used:

$$\begin{aligned} &\psi(x,t)\big|_{t=0} = \psi_k(\cos\lambda_k\xi + i\sin\lambda_k\xi), \\ &\Theta(x,t)\big|_{t=0} = -\frac{\psi_k^2}{x}, \quad \frac{\partial}{\partial t}\Theta(x,t)\Big|_{t=0} = 0, \end{aligned} \quad (14)$$

where $\Psi_k$ and $\lambda_k$ are the eigenfunction of problem (12) with the number $k$ of nodes and the corresponding eigenvalue, energy integral (11) is time-independent, i.e., is a constant coinciding with $\lambda_k$ in absolute value. Here, $\xi$ is an arbitrary factor, which we assumed to be $\pi/4$ in our calculations

Therefore, to verify the correctness of the calculation scheme, we obtained three solutions for problem (12) with the number of nodes $k = 0, 1, 2$ using the continuous analogue of the Newton method [10]. The corresponding eigenvalues are

$$\lambda_0 = 3625.55, \quad \lambda_1 = 685.97, \quad \lambda_2 = 279.01.$$

Using these solutions, we modeled the initial conditions (14) and carried out test calculations, as a result of which we established that, if $h_x = 10^{-5}$ and $h_t = 10^{-7}$ are chosen, the energy integral $W(t)$ and the shapes of the curves for the solutions of system (9) remain unchanged during a physically significant time interval. This confirms the correctness of the calculation scheme and the MPI/C++ realization.

### *Comparison of W with Theoretical Estimation*

In accordance with theoretical estimates [5], the energy integral corresponding to the steady polaron state $k = 0$ is

$$W_0(\infty) = -0.163\frac{me^4}{\hbar^2}c^2,$$

where $c = 0.552$, $m = 2.692m_e$, and $m_e$ is the electron mass in vacuum. If the fact that $E = e^2/a = 27.2$ eV, where $a = \hbar^2/(m_e e^2)$ is the Bohr radius, is taken into account, the theoretical estimation yields $W_0 = -3.637$ eV. In accordance with our calculations, after passing to dimensional units (eV), we obtain

$$\begin{aligned} W_0 &= (W_0^{num}/r_{00}^2)E \\ &= (3625.55/164.64^2) \times 27.2 = -3.638. \end{aligned} \quad (16)$$

Consequently, the values of $W_0$ calculated within the framework of our model agree with the theory.

### *Calculation of the Absorption Intensity*

To reproduce the experimental values of the intensity of light absorption by a hydrated electron, we chose the initial conditions of the wave function $\psi$ in the Gaussian form. The initial condition for the function $\Theta$ is analogous to (14). The initial value of $\varphi$ is solution of the second equation of system (9) with $\Theta(x, 0)$ on the right-hand side.

If the spherical symmetry and rescaling are taken into account, the function $\psi$ at the time instant $t = 0$ is calculated as follows:

$$\psi(x, 0) = F_g(\tilde{x})\sqrt{4\pi}\tilde{x}\sqrt{r_{00}}, \quad \tilde{x} = xr_{00}, \quad (17)$$

where

$$F_g(x) = \left(\frac{2}{\pi}\right)^{3/4}\frac{1}{\sigma^{3/2}}\exp(-x^2/\sigma^2). \quad (18)$$

The parameter $\sigma$ was chosen to obtain, in the computer simulation, the time dependence of the intensity of light absorption by water which agrees with the data in [11]. The expression used to calculate the intensity has the form [5]

$$I(\Omega, t) = \frac{4\Omega^2\gamma_s^2}{(W(t)^2 - \Omega^2)^2 + 4\Omega^2\gamma_s^2}, \quad (19)$$

where $\gamma_s = 0.38$ eV is the width of the absorption band of the hydrated electron and $\Omega$ has the physical meaning of the light frequency of a scanning laser at which light is absorbed by the hydrated electron.

The calculations were carried out for $\Omega = 1.984$ and 1.512 eV. For these two cases, Figs. 1 and 2 show the results of comparing the experimental data in [11], the approximate theoretical estimates in [5], and our obtained calculated curves. The calculated data are represented by the solid curve; the theoretical estimates, by the dashed one; and the experimental, by the saw-edged one. It can be seen that the model considered in our paper provides agreement with both the experimental and theoretical estimates on the whole. The discrepancy between numerical and theoretical estimates and experimental data in the initial graph region can be explained by the fact that formula (19) does not take the time dependence of the quantity $\gamma_s$ into account. Refinement of the method for calculating $I(\Omega, t)$ is the object of further studies.

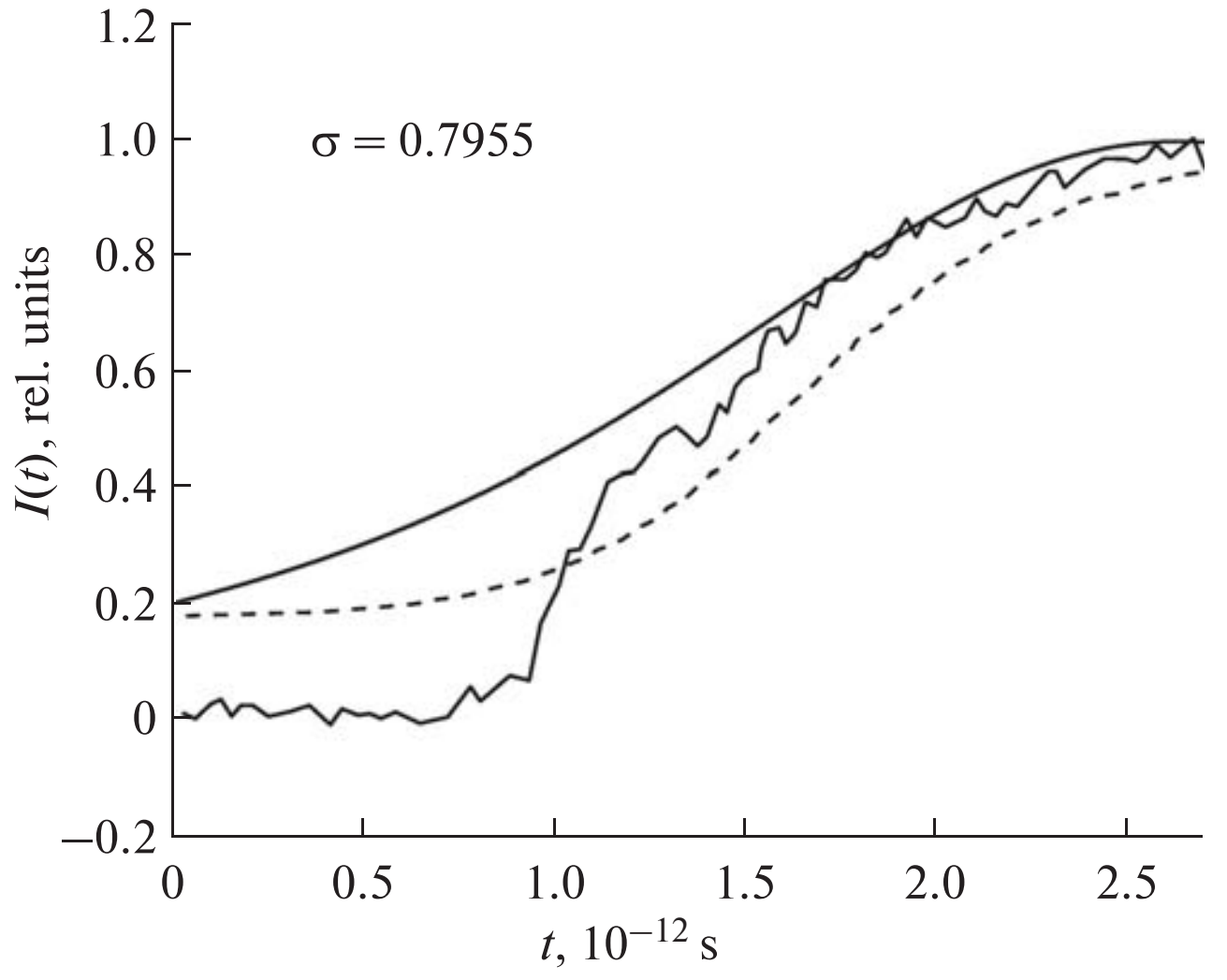

**Fig. 1.** Evolution of the intensity of light absorption by a hydrated electron for $\Omega = 1.984$ eV. The saw-edged curve corresponds to the experiment [11]; the solid smooth curve, to our calculation; and the dashed curve, to the approximate theoretical estimation [5].

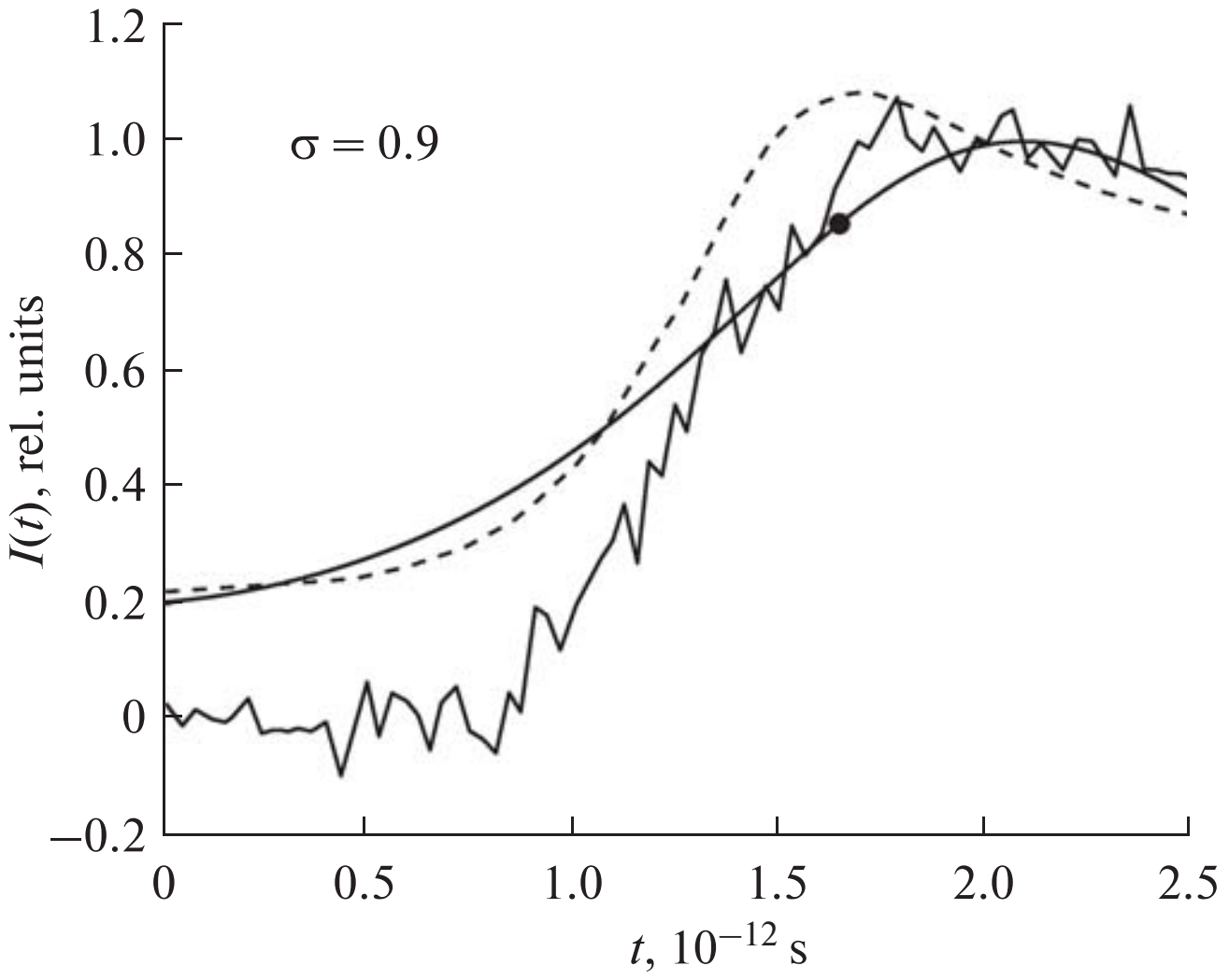

**Fig. 2.** Same as in Fig. 1, but for $\Omega = 1.512$ eV. The graph $|\psi(x)|$ shown in Fig. 4 corresponds to the bold point.

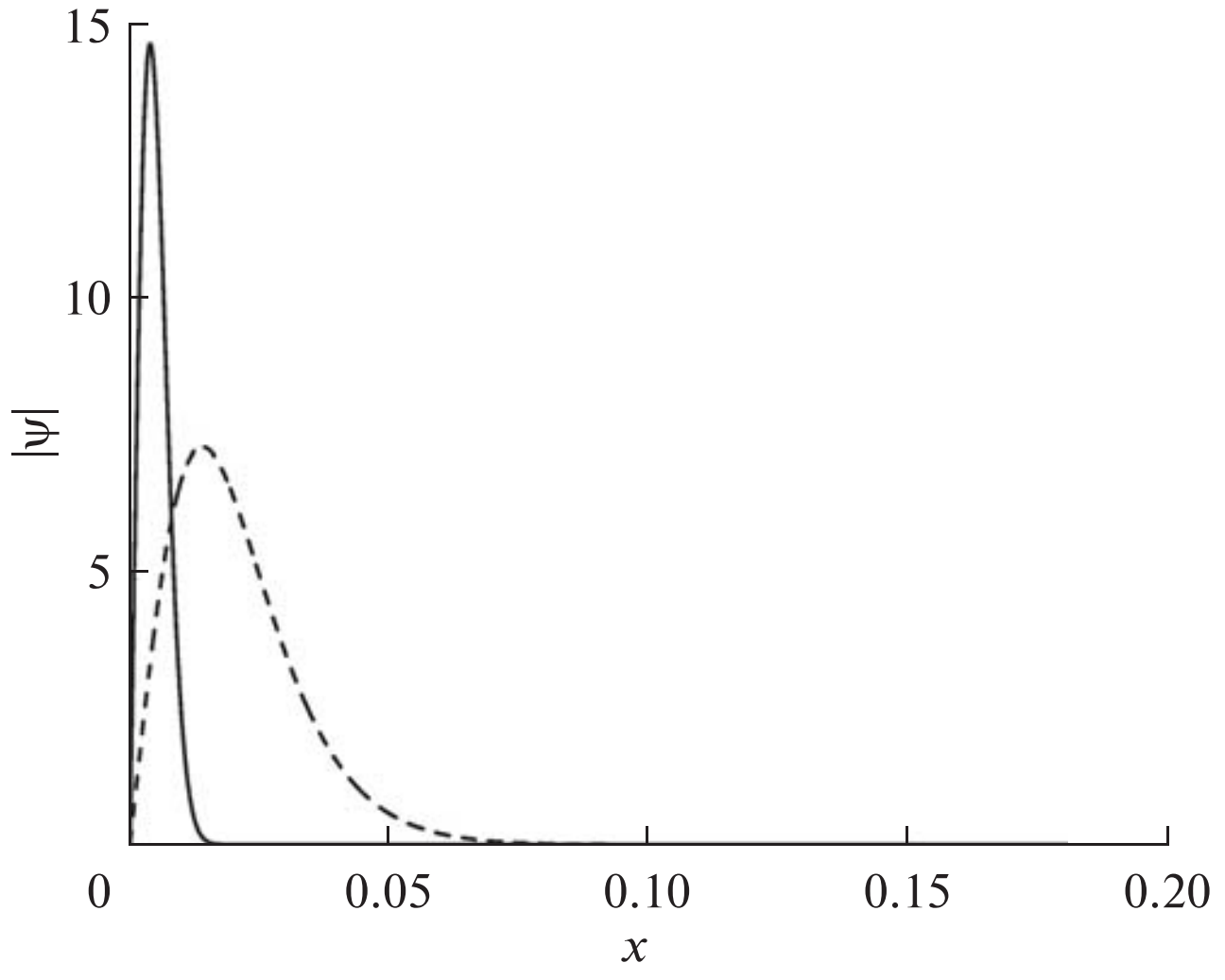

**Fig. 3.** Function $|\psi|$ at the instant of time $t = 0$ calculated using formula (17) for $\sigma = 0.9$ (the solid line) compared with the nodeless solution $\Psi_0$ of steady-state problem (12) (the dashed line).

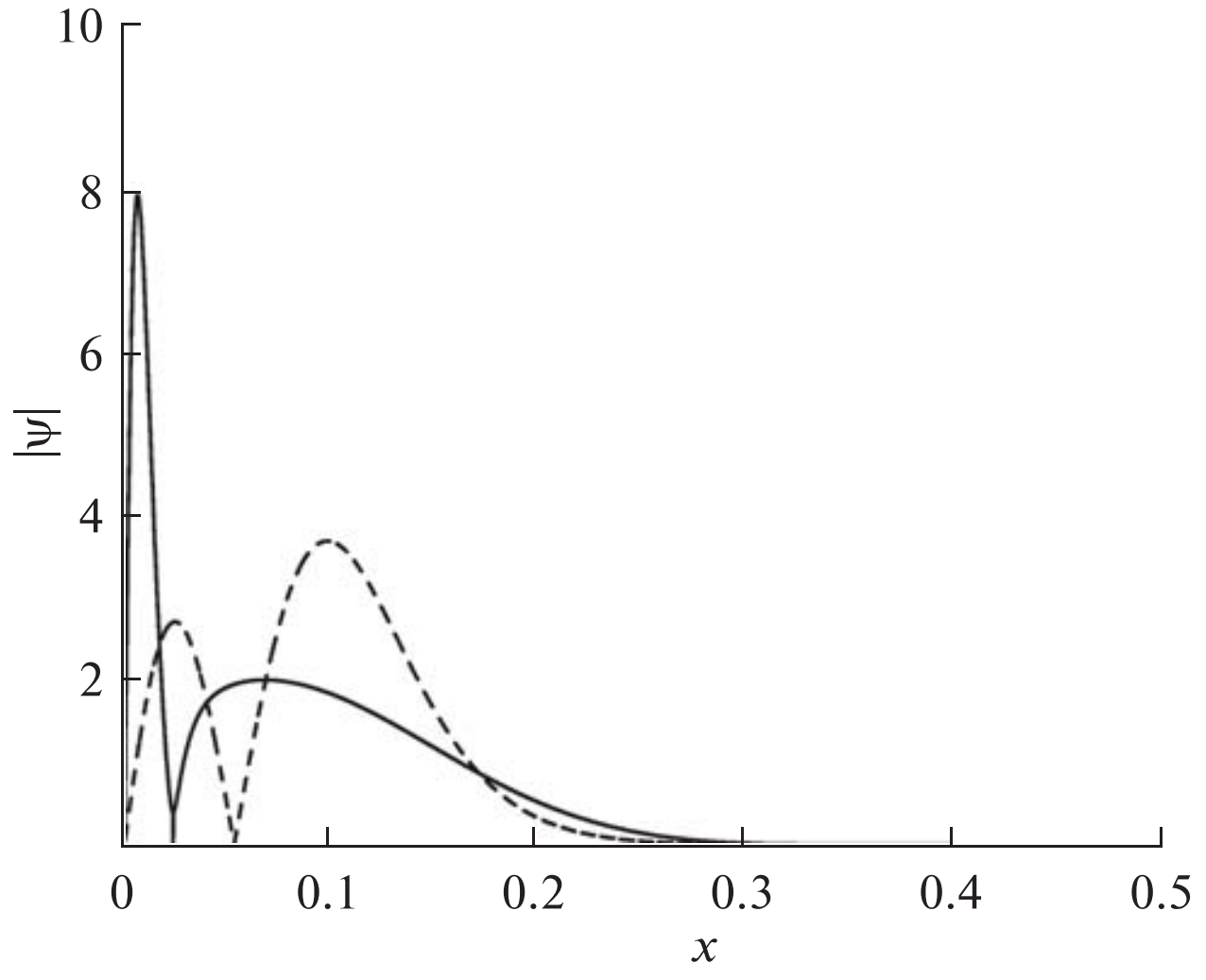

**Fig. 4.** Function $|\psi|$ for the case of $\Omega = 1.512$ eV and $\sigma = 0.9$; it corresponds to the bold point in Fig. 2 (the solid line) and is compared with the one-node solution $\Psi_1$ of the steady-state problem (12) (the dashed line).

Figures 3 and 4 demonstrate, how the wave function $\psi$ transforms during the process of computer simulation for the case shown in Fig. 2 ($\Omega = 1.512$ eV). In Fig. 3, the solid curve denotes the function $|\psi|$ at the initial time instant $t = 0$ which was calculated using formula (17). For comparison, the nodeless solution $\Psi_0$ of steady-state problem (12) is shown by the dashed curve. At times in the region of the maximum of $I(t)$, the form of the function $\psi$ corresponds to that of the one-node function. This can be seen in Fig. 4, where the solid curve shows the function $|\psi|$ corresponding to the bold point in the intensity graph in Fig. 2. For comparison, the one-node steady-state solution $|\Psi_1|$ is shown by the dashed curve. Upon further simulation, $\psi$ again acquires the form of the nodeless function as $I \to 0$. In the case corresponding to that in Fig. 1 ($\Omega = 1.984$ eV), the function $\psi$ undergoes similar transformations.

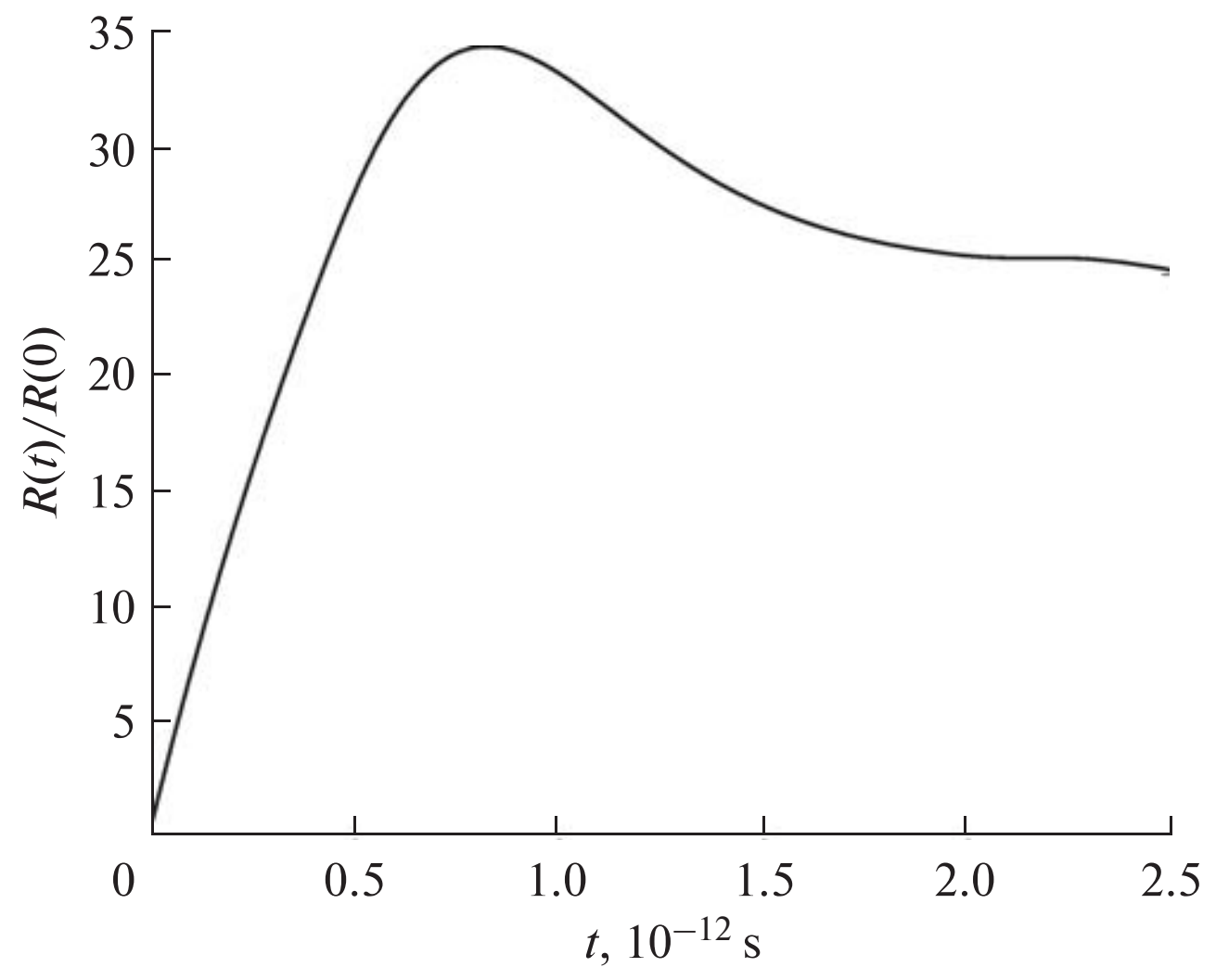

**Fig. 5.** Dependence $R(t)/R(0)$ for the case of $\Omega = 1.984$ eV and $\sigma = 0.7955$.

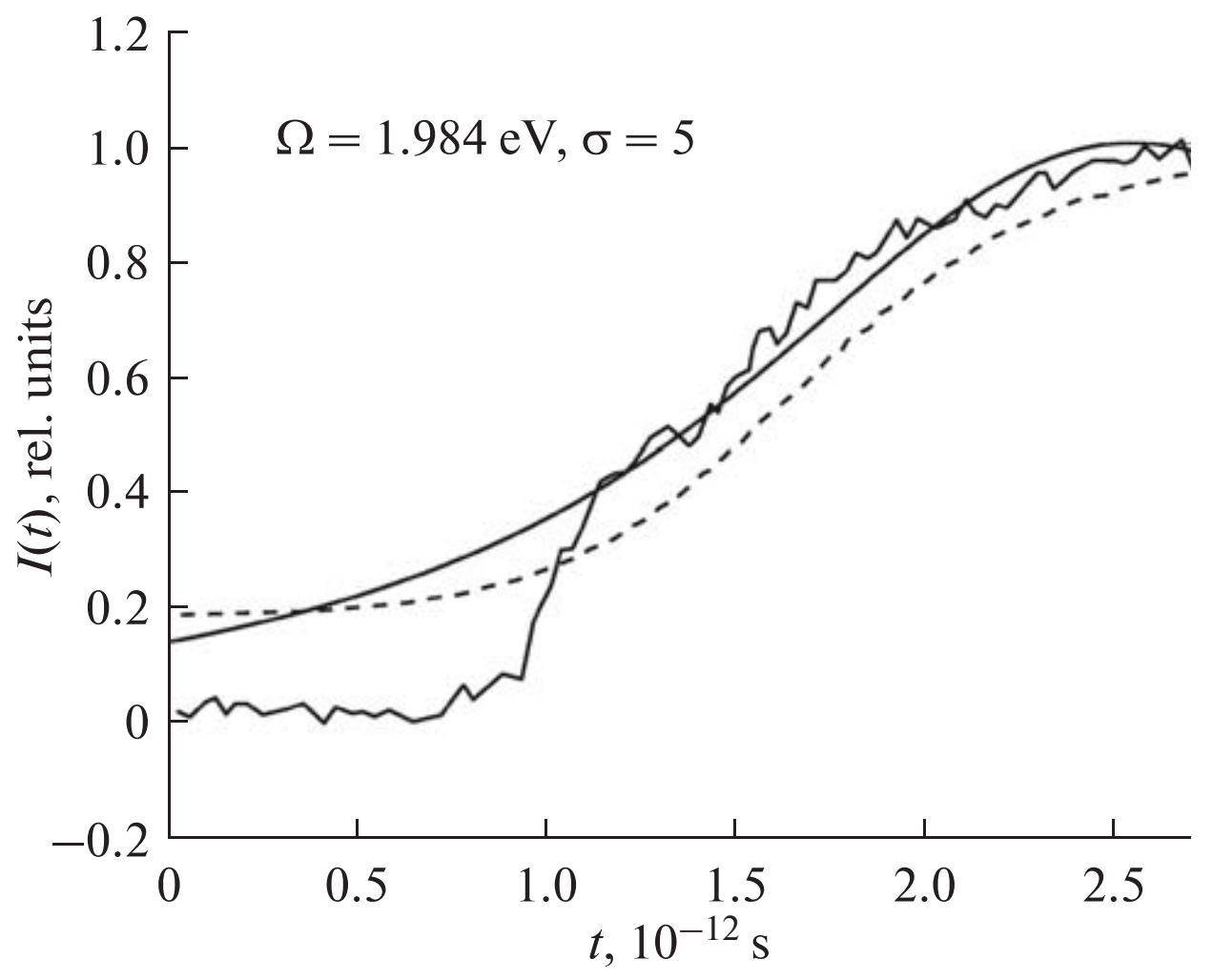

**Fig. 6.** The same as in Fig. 1, but for $\sigma = 5$.

*Calculation of the Radius of the Wave Function*

The dynamics of variation in the characteristic dimension (radius) of the wave function $\Psi$ is determined by the integral

$$R(t) = \int |\Psi|^2 r dr^3. \quad (20)$$

In terms of system (9), the formula for calculating the radius in dimensional quantities becomes

$$R(t) = r_0 \int |\psi|^2 r dr. \quad (21)$$

In Fig. 5, this dependence is shown for the case of $\sigma = 0.7955$ eV and $\Omega = 1.984$ eV. For obviousness, we here show the normalized function $R(t)/R(0)$. The radius at the initial instant of time is $0.33 \times 10^{-8}$ cm. During the evolution process, the radius $R(t)$ increases thirtyfold or more and reaches the maximum earlier than the function $I(t)$ in Fig. 1. Then the radius decreases gradually and stabilizes at values near $8.3 \times 10^{-8}$ cm in the region of the maximum of the function $I(t)$.

The final value of the radius does not contradict well-known estimates of (2–3) $\times 10^{-8}$ cm [1]. However, from physical considerations, other dynamics of radius variation during polaron-state formation should be expected, namely, variations from larger values at the initial instant of time to lower ones. Therefore, we carried out additional calculations to seek alternative values of $\sigma$ that adequately reproduce the dynamics of $I(t)$ observed in the experiments and provide a decrease in the radius in this case. Simultaneous fulfillment of these requirements is reached for cases of $\Omega = 1.984$ and $1.512$ eV for $\sigma = 5$ and $5.75$, respectively. The corresponding intensity curves (compared with the theoretical and experimental results) are given in Figs. 6 and 7. The dynamics of the radius variation (in cm) is shown in Fig. 8 by the solid curve for $\Omega = 1.984$ V and by the dashed one, for $\sigma = 5$. It can be seen that the value of the radius in this case decreases with increasing $t$; however, an initial value of (2.4–2.6) $\times 10^{-8}$ cm and a final one of (1.5–1.7) $\times 10^{-8}$ cm are close to the above-mentioned estimate of (2–3) $\times 10^{-8}$ cm. Thus, the approach developed above makes it possible to adequately reproduce the dynamics of variation in the intensity of light absorption by a hydrated electron.

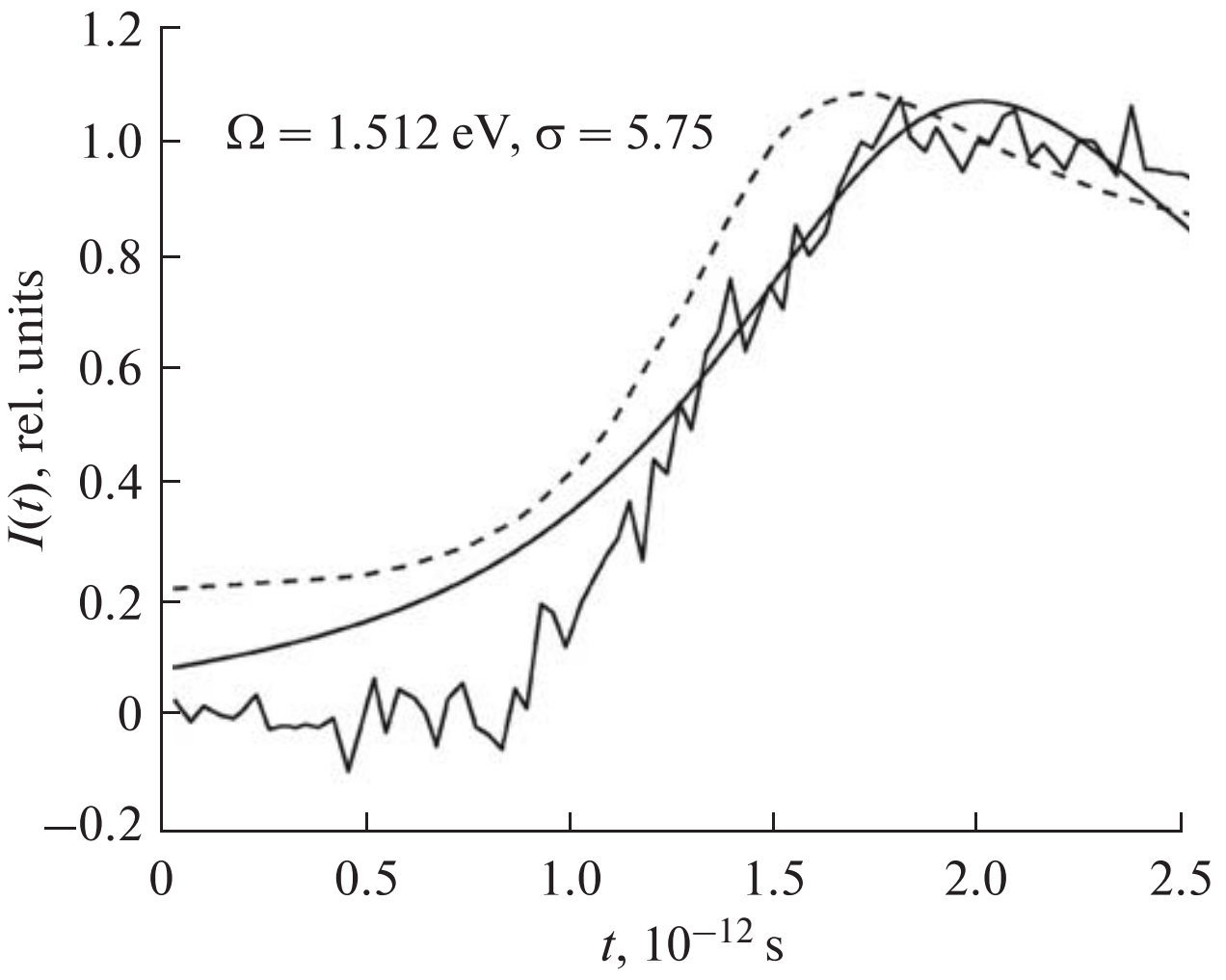

**Fig. 7.** The same as in Fig. 2, but for $\sigma = 5.75$.

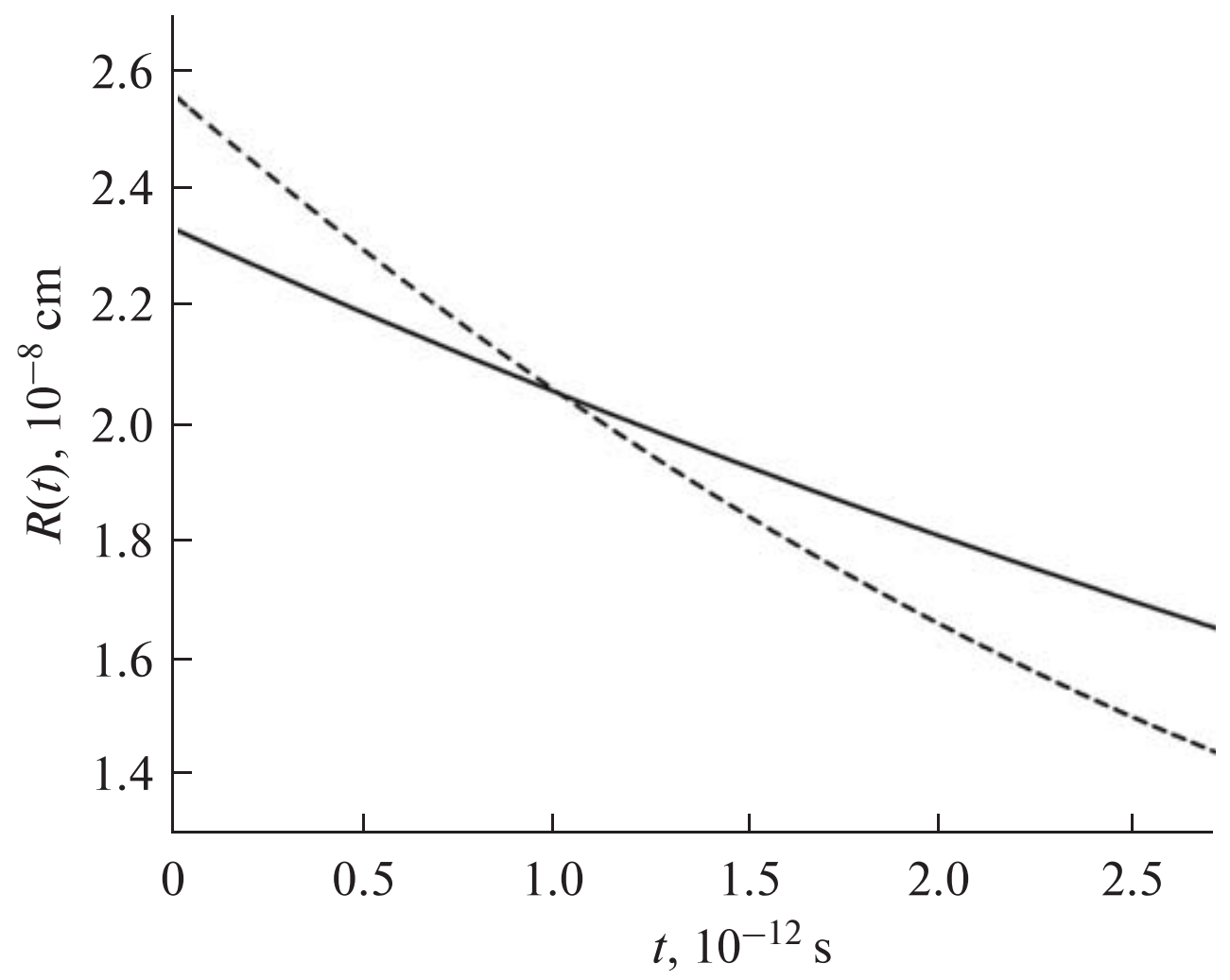

**Fig. 8.** Dependence $R(t)$ for the case of $\Omega = 1.984$ eV and $\sigma = 5$ (the solid curve) and for the case of $\Omega = 1.512$ eV and $\sigma = 5.75$ (the dashed curve).

## CONCLUSIONS

The results of numerical simulation lead to interesting conclusions. First of all, they show that the initial state of the photoexcited "dry" electron is rather localized: in accordance with Fig. 8, its radius $R(0)$ is $\sim 2.5 \times 10^{-8}$ cm. Because electron photoabsorption occurs to the water "conduction band" in accordance with the model under consideration, we can conclude that, even in pure water, this band contains potential fluctuations (for example, the cavity), which capture photoexcited electrons, forming initially bare localized states. During the process of further evolution, the fluctuation potential wells become deeper, which leads to a decrease in the electron energy and in its radius $R(0) \sim 1.5 \times 10^{-8}$ cm. On the whole, such a picture corresponds well to modern ideas of the dynamics of hydrated electrons.

Thus, we have developed and realized via programs the dynamic model of the formation of hydrated-electron states. The correctness of the operation of the corresponding computer program was confirmed by comparing the calculated results with the theoretical estimate of the energy integral and on the basis of simulation with the initial conditions in the form of solution of the steady-state problem.

We have shown that, within the framework of the considered approach, it is possible to adequately reproduce the results of the experiment on the formation of photoexcited electrons in water under laser irradiation in the ultraviolet range. The model can be used in further calculations and predictions when studying the dynamics of polaron states in the water medium and other condensed media.

## ACKNOWLEDGMENTS

The work was supported in part by the Russian Foundation for Basic Research (grant nos. 12-01-00396, 13-01-00595, and 13-01-00060).

*Translated by L. Kul'man*